\documentclass[12pt]{article}
\def\be{\begin{equation}}
\def\ee{\end{equation}}
\def\la{\label}
\def\bea{\begin{eqnarray}}
\def\eea{\end{eqnarray}}
\def\non{\nonumber}
\def\ci{\cite}
\def\la{\label}
\def\bib{\bibitem}
\def\lesssim{{_ <\atop{^\sim}}}
\def\lm{\lambda}
\def\Lm{\Lambda}
\def\le{\left}
\def\ri{\right}

\def\rb{\rho_b}
\def\rp{\rho_\phi}

\def\gm{\gamma}
\def\gmgm{\gamma_b}
\def\gmb{\gamma_b}
\def\gmp{\gamma_\phi}
\def\gmptr{\gamma_{\phi tr}}
\def\gmpo{\gamma_{\phi o}}

\def\D{\Delta}
\def\Omp{\Omega_\phi}
\def\Ompi{\Omega_{\phi i}}

\def\Om{\Omega}
\def\wtr{w_{tr}}
\def\wptr{w_{\phi tr}}
\def\wp{w_\phi}

\def\Ompo{\Omega_{\phi o}}
\def\wpo{w_{\phi o}}

\def\s8{\sigma_8}

\def\fr{\frac}

\def\raw{\rightarrow}

\usepackage{graphicx}

\begin{document}

\begin{flushright}
astro-ph/0211519

\end{flushright}

\begin{center}
   {\Large \bf
           Implications to CMB from Model Independent evolution of $\wp$
            and Late Time Phase Transition
   }

\end{center}

\vspace*{0.05cm}

\begin{center}
{\bf A. de la Macorra\footnote{e-mail: macorra@fisica.unam.mx}
  }
\end{center}

\vspace*{0.05cm}

\begin{center}
{\small
\begin{tabular}{c}
Instituto de F\'{\i}sica, UNAM\\
Apdo. Postal 20-364, 01000  M\'exico D.F., M\'exico\\
\\
\end{tabular}
}
\end{center}

\vspace{.1 cm}

\begin{center}
{\bf ABSTRACT}
\end{center}
\small{
We present model independent determination of the CMB  from
any kind of fluid that has an equation of state
taking four different values. The first region has
$w=1/3$, the second $w=1$, the third $w=-1$
while the last one has $-1<w=cte<-2/3$.
This kind of dynamical $w$  contains as a limit
the cosmological constant and tracker models.

We  derive the model independent evolution of $\wp$, for
scalar fields, and we
see that it remains most of the  time in either of its three extremal
values given by $\wp=1,-1,w_{tr}$. This "varying" $w$
is the generic behavior of scalar fields, quintessence, and we determine
the size of the different regions by solving the dynamical equations
in a model independent way.

The  dynamical $w$ models have a better fit to CMB data then the
cosmological constant  and the tracker models. We determine
the effect of having the first two regions $w=1/3,1$ and depending
on the size of these periods they can be observed in the CMB.

These models can be thought as arising after a late time
phase transition where the scalar potential is produced.
Before this time all the fields in this sector were massless and
redshifted as radiation, giving the first period $w=1/3$.

In general, the CMB spectrum sets a lower limit
to $\D N_T$ and to the phase transition scale $\Lm_c$.
For smaller
$\D N_T$ the CMB peaks are moved to the right of the spectrum
and the hight increases considerably.

Depending on    the initial energy density we
obtain a lower limit to the phase transition scale $\Lm_c$,
when the scalar field appears and we have the transition
from $w=1/3$ to $w=1$. For
 $\Ompi=0.1$ the CMB sets a  lower
limit to the phase transition scale  $\Lm_c\geq 0.2 eV$.
For inverse power low potentials with $\Ompi\leq 0.1$
the constrain $w_o<-2/3$ requires
a power $n\leq 1.8$ and a phase transition $\Lm_c\geq 4 MeV$
leaving a small energy scale  window for models to work.

  }


\noindent \rule[.1in]{14.5cm}{.002in}

\thispagestyle{empty}

\setcounter{page}{0} \vfill\eject

\section{INTRODUCTION}

In recent time the cosmological observations on
the  cosmic microwave background radiation ("CMB")  \ci{CMB} and
the supernova  project SN1a \ci{SN1a}
have lead to conclude that the
universe is flat and it is expanding with an accelerating
velocity.  These conclusions show that the universe is now
dominated by a energy density with negative pressure with
$\Om_{\phi}=0.7 \pm 0.1$ and $ w_\phi < -2/3$ \ci{w}.
This energy
is generically called the dark energy.
Structure
formation also favors a non-vanishing dark energy
  \ci{structure}.

It is not clear yet what this dark energy is.
It could be a cosmological constant, quintessence
(scalar field with gravitationally interaction) \ci{tracker}
or some other kind of exotic energy.

The best way to determine  what kind of energy
is the dark energy is trough the equation of
state parameter $\wp=\gmp-1=p/\rho$, where
$p$ is the pressure and $\rho$ the energy density of the fluid,
and through its imprint  on the CMB.
The survey of redshifts of the different objects should
in principle allow as to determine the value of
$\wpo$ ($o$ subscript referees to present day quantities)
but only at  small redshifts $z$. The result from the SN1A
project \ci{SN1a} sets un upper limit to $\wpo<-2/3$ but does not
distinguish a cosmological constant with (constant)
$\wp\equiv-1$ and quintessence
 or any other form of matter with $\wpo<-2/3$. It would
 be very interesting if in the future the SN1a survey could
 constrain better the value of $\wpo$.

The CMB could give us information not only on  the value
of $\wpo$ but also on its form during all matter domination era.
We will study models that have a changing $\wp$ over
time, well defined by the dynamics.
We would like to see if we can distinguish from  the CMB spectrum
between a varying $\wp$, a fixed $\wp=cte$ and  a true cosmological constant.
Some  general approaches can be found in \ci{gen.dark}.

We will analyze the contribution to the CMB from a
dark energy with a $\gmp=\wp+1$ that takes four
different values. It will have a $\wp=1/3$
for energies above a certain
scale $\Lm_c$, which we will call the phase transition scale.
Starting at $\Lm_c$  we will have  a region with $\wp=1$
and duration  $\Delta N_1$,
where $N$ is the logarithm  of the scale factor $a$ ($N=Log[a]$).
Thirdly we will have $\wp=-1$ for almost the same amount
of time as in the previous  period, $\Delta N_2 \simeq \Delta N_1$,
and finally we will end up in a region with $-1\leq \wpo=cte \leq -2/3$
for a duration of $\Delta N_o$.
The cosmological evolution and the resulting CMB will have only four
new parameters $\Delta N_1, \Delta N_2, \Delta N_o$ and $\wpo$. By varying
these parameters we will cover a wide range of models. In
particular we will cover all quintessence models.

The analysis of the CMB with this
kind of dark energy does not depend on its nature,
 it could be a scalar field (quintessence)
or any other form of dark energy that gives the four sectors described
above. However, we would like to point out that this
pattern of different $\wp$ is precisely what one expects from
a quintessence scalar field and we will prove it. We will
also show that it is generic, i.e. model independent.
In the case of a quintessence scalar field the parameters
 $\Delta N_1,\Delta N_2, \Delta N_o, \wpo$ will be functions of
 $\Lm_c$, the phase transitions scale where the scalar field
 is produced, $\Ompi$ the initial energy density of the scalar field,
 the minimum value of $y_{min}\equiv V(\phi)/3H^2$
and on the final value $\wpo$. For inverse power low
potentials "IPL"  the number of parameters is reduced
to three, $\Ompi, \Lm_c, \wpo$.

The evolution of scalar fields has been widely
studied and some general approaches con be found in \ci{generic,mio.scalar}.
The evolution of the scalar field $\phi$ depends
on the functional form of its potential $V(\phi)$ and a late time
accelerating universe constrains the form of the potential
\ci{mio.scalar}. Even though the evolution
of the scalar field depends on the potential we
will show that it is possible to obtain a model
independent behavior of $\Omp$ and $\wp$.

The contribution of scalar fields
to the CMB has been studied but in most
cases a constant $\wp$ has been used. The fields
with constant $\wp$ are called tracker fields \ci{tracker}.
Even though tracker fields are very interesting, specially because
they do hardly depend on the initial conditions, they
are not consistent with the observed $\wp$ (at least
for inverse power potentials) .
This work generalizes the tracker analysis since it contains
the tracker model as a limiting case

IPL tracker fields with constant $\wp=\wptr=cte$ are not
 consistent with present day cosmological  observations.
Tracker fields require $N>5$ and
a small $\wpo<-2/3$ today requires $n<1$.  However,
 for $n<1$ the scalar field has not reached its tracker value
 by present day.
Of course, tracker fields are not the generic evolution of
scalar fields.

We will show that the  generic behavior of $\gmp=\wp+1$
for a quintessence scalar field with an arbitrary potential
(with the restriction $V\geq 0$ and  $\lm_i=-V'/V \gg 1$)
has three critical points given by  $\gmp=2,0$ and $\lm^2\Omp/3$ (or
$\wp=1,-1$ and $\lm^2\Omp/3-1$).

The parameter $\gmp$ will be
most of the time in either of the three critical points.
Independent of its initial value it will go rapidly
to $\gmp=2$ and remain there for a long period of time
$\Delta N_1$.
Afterwards  it will sharply go
to $\gmp=0$ and stay there during almost the same amount of time
as in the first stage $\Delta N_2\simeq\Delta N_1$. The
amount of time it spends in these two regions
depends only on $\Ompi$, the initial energy density,
and on $\Lm_c$ the phase transition scale.

Finally, $\gmp$  will evolve  to its
tracker value
$\gmptr=\lm^2\Omp/3$ where it will remain. The
amount of time before we reach present day, denoted
by $\Delta N_o$, depends
on the values of  $\Ompi$ and $\gmptr$.

Using this generic evolution of $\gmp$ we can
determine which models have the best fit to
the acoustic CMB pattern by varying
$\Delta N_1, \Delta N_2$, $\Delta N_o$ and $\gmpo$.
The change of these four parameters
covers all scalar field models.
We hope to
be able to infer form the results the phase transition scale
$\Lm_c$.

The work is organized as follows. In sect.\ref{over} we give
an overview of the models and we summarize the main theoretical and
phenomenological results. In sect.\ref{evol} we set the general dynamical
 equations for the quintessence field. In sect.\ref{indep} we first analyze
 in a model independent way the evolution of $\phi$ and then we do
 a model independent analysis of the dynamics of  $\gmp=\wp+1$. In both cases we
 give the model dependent parameters.  In sect.\ref{duration} we show
 how long the regions with $\gmp=2,0,\gmpo$ last while in sect.\ref{anal}
 we compare the CMB obtained in the presence of the scalar field
with the experimental data and
 with a true cosmological constant. Finally we give in sect.\ref{concl}
 our conclusions.

If the reader is only interested
on the setting of the model and its cosmological consequences
she/he could skip sections \ref{evol} and \ref{indep} where
we prove the model independence behavior of the $\gmp=\wp+1$
and go directly to sect.(\ref{duration}).

\section{Overview}\la{over}

The strategy is to analyze the spectra of CMB, using
 a modified version of CAMB,
in a model independent
way and see from its result if we can distinguish between different
quintessence models, tracker,  cosmological constant or other kinds
of exotic energy densities. The results
on the effect on CMB by the  fluid with a generic
behavior of $\gmp=\wp+1$, as seen from fig.(\ref{fig1}) will be valid
independently of the nature of this fluid, i.e. scalar field
or a exotic type of fluid. Notice that the $w=1/3,1,-1,w_{tr}=cte$
model fits well with the numerical result of a IPL potential
with $n=1$ and $\Ompi=0.05$.

In the case of a scalar field,
we will assume that the scalar field appears at a scale $\Lm_c$
with an energy density $\Omp(\Lm_c)$. The late time appearance
of the $\phi$ field suggests that a phase transition takes
place creating the scalar field. We are not concern
with the precise mechanism of its
appearance (see \ci{chris1,ax.mod}). However, energy conservation would suggest
that the energy density of the $\phi$ field after the phase
transition would be given in terms of
the energy density of the system  before the phase transition
and we will take them to be equal.
It is natural to assume that all the energy density
before the phase transition, in this sector, was
in relativistic degrees of freedom. If the phase transition
takes place after nucleosynthesis "NS"  then the primordial
creation of nuclei puts un upper limit to the relativistic energy
density to be less than 0.1-0.2 of the critical energy density \ci{NS,NSC}.
If $\Lm_c$ is larger than the NS scale then we do not need
to worry about the NS bound since independent of its initial
value,  $\Omp$ will drop rapidly and remain small for
a long period of time (covering NS).

In a chronological order, we would start
with a universe filled with the SM particles and a Q
sector (could be another gauge group) and with gravitational
interaction between the two sectors only.
In both sectors
all fields start massless, i.e. they redshift
as radiation. The evolution
of the SM is the standard one and we have nothing new
to say. However, the Q sector will have a
phase transition at $\Lm_c$ leading to the appearance
of a scalar field $\phi$ with a potential $V(\phi)$, the quintessence field.
Above $\Lm_c$ the fields in this
sector will behave as radiation.
The evolution of $\phi$ for energies below $\Lm_c$
is that of a scalar field with given potential $V$.
However, the precise form of $V$ is unknown.
In table \ref{tab1} we show the different model independent regions
that we consider. The model dependence lies only on the
size of these different periods and on the value
of $\gmp$ in the last region.

If a late time phase transition takes place, so
that most of the time the universe has been dominated by
matter, then $\Delta N_1\simeq \Delta N_2$ as seen from eq.(\ref{N1}).
This will be the case for a transition scale $\Lm_c$ smaller
than the radiation-matter equality energy $"E_{rm}"$. If
$\Lm_c \gg E_{rm}$ we
have a large radiation domination epoch
and then $\Delta N_1\simeq 2\Delta N_2$.

From a cosmological point of view
we have only 4  free parameters $\Delta N_1,\Delta N_2$, $\Delta N_o$
and $\gmptr$ (the value of $\gmp$ during the third period).
With these  parameters we cover all models.

\begin{table}
\begin{center}\la{tab1}
 \begin{tabular}{|c|c|c|c|}
    \hline
   Sector & Energy & Duration & $\gmp=\wp+1$ \\ \hline
   Radiation & $E> \Lm_c$  & $N>N_i$ & 4/3 \\
   First & $E_1< \Lm_c$& $N_i<N<N_1$ & 2 \\
   Second& $E_2< E_1$ & $N_1<N<N_2$ & 0 \\
   Third & $E_3< E_2$ & $N_o<N<N_2$ & $\lm^2\Omp/3$ \\
   \hline
 \end{tabular}
 \caption{\small{We show the different regions, its duration
 and the value of $\gmp$ in each region with $N=Log[a]$.
 With $\Delta N_1\equiv N_1-N_i$ and $\Delta N_2\equiv N_2-N_1$
 one obtains $\Delta N_1=d\Delta N_2$, d=1,2 for
matter  or radiation dominance, and  all model dependence
is then given by $\Delta N_2, \Delta N_o=N_o-N_2$ and $\gmp$ in the
third region ($N_o$ is at present day). }}
\end{center}
\end{table}

\begin{figure}[htp!]
\begin{center}
\includegraphics[width=14cm]{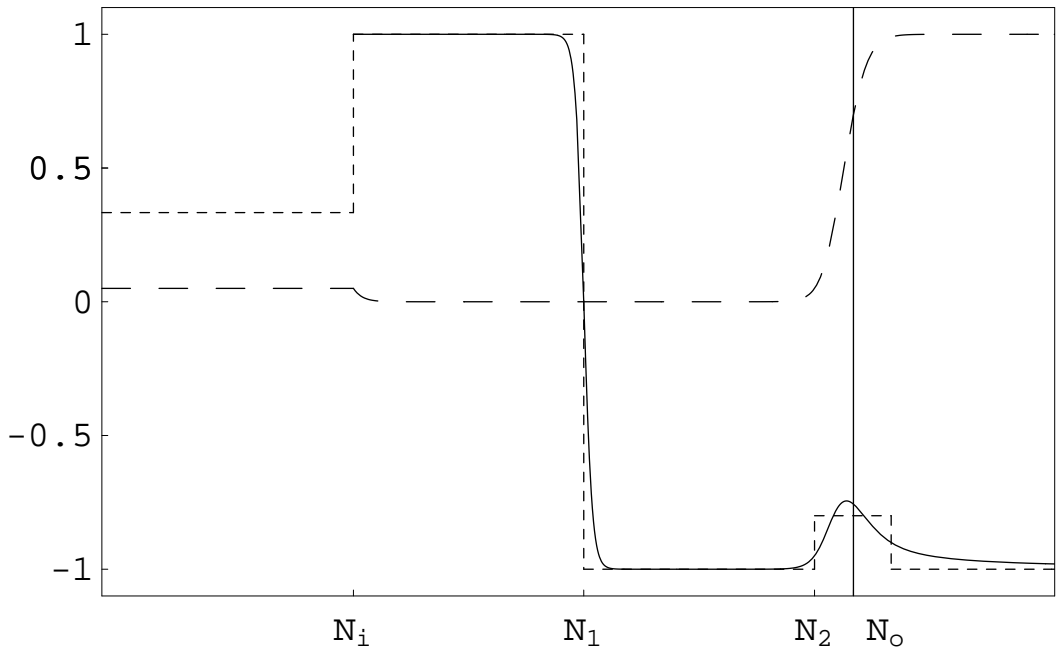}
\end{center}
\caption{\small{We show the evolution of $\wp$ and
$\Omp$, solid and dashed lines respectively for
an IPL potential with $n=1$ and $\Ompi=0.05$
 in a matter background as a function of $N=Log[a]$.
The dotted line represents the theoretical $\wp$
and we see that it makes a good fit to the numerical
solution. $N_i$ is given at the initial
scale $\Lm_c$ and $N_1,N_2$ give the end of the regions
with $\wp=-1,1$ respectively while the solid vertical
line at $N_o$ denotes
present day. Notice that for $N<N_i$ we are assuming
that the energy density $\rp$ redshifts as radiation and
we are also assuming that radiation dominates for $N<N_i$.
}}
\la{fig1}
\end{figure}

The   cosmological parameters are given in terms of
the field theoretical  model dependent parameters.

\section{Cosmological Evolution of $\phi$}\la{evol}

We will now determine the cosmological evolution of a scalar field
 $\phi$  with arbitrary potential $V(\phi)$ and with only gravitational
 interaction with all other  fields. This field is called
 quintessence.

The cosmological evolution of $\phi$ with an arbitrary potential
$V(\phi)$ can be determined from a system of differential
equations describing a spatially flat Friedmann--Robertson--Walker
universe in the presence of a barotropic fluid energy density
$\rho_{b}$ that can be either radiation or matter. The
equations are
\bea\la{eqFRW}
\dot H &=& -\frac{1}{2}(\rho_b+ p_b+\dot \phi^2),\non \\
\dot \rho &=& -3H(\rho+p),\\
\ddot \phi &=& -3H \dot \phi-\frac{d V(\phi)}{d \phi},\non
 \eea
where $H$ is the Hubble parameter, $\dot \phi = d\phi/d t$, $\rho$
($p$) is the total energy density (pressure).  We use the change
of variables $x \equiv \frac{\dot \phi}{\sqrt{6} H}$ and $y \equiv
\frac{\sqrt{V}}{\sqrt{3} H }$
   and  equations (\ref{eqFRW}) take the following form
  \cite{liddle,mio.scalar}:
 \bea  \la{eqFRW2}
x_N&=& -3 x +\sqrt {3 \over 2} \lambda\,  y^2 + {3 \over 2} x
[2x^2 + \gm_{b} (1 - x^2- y^2)] \non\\
 y_N&=& - \sqrt {3 \over
2} \lambda \, x\, y + {3 \over 2} y [2x^2 + \gm_{b} (1 - x^2
-y^2)]\\
H_N&=& -{3 \over 2} H [2x^2 + \gm_{b} (1 - x^2 - y^2)] \non
 \eea
where $N$ is the logarithm of the scale factor $a$, $N\equiv
ln(a)$; $f_N\equiv d f/d N$ for $f=x,y,H$; $\gm_b= 1+ w_b $
and $\lambda(N)\equiv -V'/V$ with $V'= dV/d\phi$. In terms of $x,
y$ the energy density parameter is $\Om_{\phi}=x^2+y^2$ while the
equation of state parameter is given by $\gm_\phi-1=w_{\phi}\equiv
p_{\phi}/\rho_{\phi}=\frac{x^2-y^2}{x^2+y^2}$. It is clear
that $x^2, y^2 \leq 1$.

The Friedmann or constraint equation for a flat universe
$\Omega_{b}+\Om_{\phi}=1$ must supplement equations
(\ref{eqFRW2}) which are valid for any scalar potential as long as
the interaction between the scalar field and matter or radiation
is gravitational only. This set of differential equations is
non-linear and for most cases has no analytical solutions. A
general analysis for arbitrary potentials is performed in
\cite{generic,mio.scalar}. All model
dependence falls on two quantities: $\lambda(N)$ and the constant
parameter $\gm_b=1,4/3$ for matter or radiation, respectively.
We will be interested in studying scalar fields that lead to a
late time accelerated universe, i.e. to quintessence, and in this
case we will have a decreasing $\lambda(N)$ \ci{mio.scalar}
and a late time behavior $\lambda(N) \rightarrow 0$.
For constant  $\lambda(N)$ (exponential potential) one can have an accelerating
universe if $\lambda(N)< \sqrt{6}$ but its dynamics  would lead to
an accelerating universe too rapidly, i.e. not at a late time
as ours, unless we fine tune the initial conditions.

It is also useful to have the evolution of $\Omp=\rp/3H^2=x^2+y^2$
and $ \gmp=1+\wp=2x^2/(x^2+y^2) $, derived from eq.(\ref{eqFRW2}),
\ci{ax.coinci}
 \bea\la{dOm}
   (\Omp)_N&=& 3 (\gm_b-\gmp) \Omp (1- \Omp)\\
 (\gmp)_N &=& 3\gmp (\gmp-2)\le(\lm \sqrt{\fr{\Omp}{3\gmp}}-1 \ri)
\la{gmp}\eea

The evolution for
the energy density, valid only for constant $\gmp$, is the usual
one
\be\la{rp}
\rp=\rho_{\phi i} e^{-3(N-N_i)\gmp}.
\ee
and the evolution of $\Omp$ when it is much smaller than one and with
constant $\gmp$ is
\be\la{omp}
\Omp=\Ompi e^{-3(N-N_i)(\gmp-\gmgm)}.
\ee
From now the subscript $i$ stands for initial conditions,
when the potential $V$ appears, and
the subscript $o$ for present day values.

\section{Model Independent Analysis}\la{indep}

\subsection{Evolution of $x, y$ and $H$}\la{evol.xyH}

We are interested in studying scalar potentials
that lead to quintessence, i.e. a late time (i.e. present day)
acceleration period of the universe. For this to happen one
needs  $\lm=-m_{pl} V'/V \raw 0$ in the asymptotic limit
(or to a constant  less then one).
An accelerating universe (slow role conditions)
 requires $|\lm|<1$ and we want this period
to be at a late time.
We will consider potentials with $V\geq 0$ and since
the $\phi$ field evolves to its minimum $V'<0$ and $\lm \geq 0$ where
we are assuming,
without loss of generality, models with $\phi \geq 0$.

We will define the phase transition scale $\Lm_c$ in terms
of the potential by
\be\la{Lmc0}
\Lm_c=V_i(\phi_i)^{1/4}
\ee
where $V_i$ is the initial value of the potential
and we will consider models that have an initial value
\be\la{lmi}
 \lm_i=-m_{pl} \fr{V'(\phi_i}{V(\phi_i)} \gg 1.
\ee
From dimensional analysis we expect
$\lm_i=O(m_{pl}/\Lm_c) \gg 1 $.
If we have a phase transition at a scale $\Lm_c$ which leads to
the appearance of the $\phi$ field (e.g. composite field) then
we would also expect $\phi_i\simeq \Lm_c$. We will be working
with late time phase transition but $\Lm_c$ could
be  as large as $10^{16} GeV$ and we will still have
 $\lm_i \gg 1$.

An interesting general property
of these models is the presence of a many e-folds scaling period
in which $\lm$ is practically  constant and $\Om_{\phi} \ll 1$.

A semi-analytic approach \ci{chris2} is useful to study some properties of the
differential equation system given by eqs.(\ref{eqFRW2}). To do
this we initially consider only the terms that are proportional to
$\lm$, since $\lm_i \gg 1$, then we follow the evolution of $x$, $y$
and $H$ so every period has a characteristic set of simplified
differential equations. We see from eqs.(\ref{eqFRW2}) that the
leading terms in $x$ and $y$, for $\lm \gg 1$, are
$x_N= \sqrt {3 \over 2} \lambda\,  y^2 $ and
 $y_N= - \sqrt {3 \over 2} \lambda \, x\, y$.
Combining these equations we have
\be\la{xy1}
x_Nx=-y_Ny
\ee
with a constant circular solution
\be\la{ompi}
\Omp\equiv x^2+y^2 =x_i^2(N_i)+y_i^2(N_i)\equiv \Ompi (N_i).
\ee
Since $x_N$ is positive $x$ will grow
while  $y_N$ is negative giving a decreasing $y$. This period
ends at a scale $N_{min}$ with $x^2(N_{min})\simeq \Ompi(N_i) \gg y_{min}^2$.
Since $\lm_i \gg 1$,  the $x$ and $y$ derivatives are quite
large and the amount of
e-folds between the initial time with $y_i$ until
$y$ reaches its minimal value $y_{min}$ is very short.
An easy estimate
can be derived from $y_N/y = -c \lm \gg 1, \;c=\sqrt{3/2}x$
giving  $1\gg N_{min}-N_i=Log[y_{min}/y_i]/c\lm_i = O(1/\lm_i)$, in the
assumption $c\lm_i=cte$.

The minimal value of $y$, given  at $N_{min}$, can be obtained
from eq.(\ref{eqFRW2}) with $y_N=0$. At his point we have
\be\la{lm1}
\lm(N_{min})= -\sqrt{\fr{2}{3}}\fr{H_N}{H x}=\sqrt{\fr{3}{2}}
\fr{[\gmgm+\Ompi(2-\gmgm)]}{\sqrt{\Ompi}}\simeq \fr{1}{\sqrt{\Ompi}}
\ee
where we have taken  $x^2(N_{min})\simeq \Ompi$ and
$H_N/H=-3/2(\gmgm+\Ompi(2-\gmgm))$ since $y^2_{min} \ll 1$.
We see that $\lm$ in eq.(\ref{lm1}) is of order $1/\sqrt{\Ompi} $
and we have $\lm_i/\lm(N_{min}) \gg 1$.

The value of $y_{min}$ depends on the functional form
of $V(\phi)$, which sets the functional form of $\lm=-V'/V$.
In general we have $y_{min}=V(\phi_{min})/3/H^2_{min}$ but
without specifying $V(\phi)$ it is not possible to determine
$y_{min}$.

For an inverse power  law potential with $V=\Lm_c^{4+n}\phi^{-n}=3y^2H^2$
one has
\bea\label{ymin}
 y_{min}&=&\fr{\Lm_c^{\fr{4+n}{2}}\phi_{min}^{-n/2}}{\sqrt{3} H_{min}}\\ \non
&=&y_i\le(\fr{\phi_i}{\phi_{min}}\ri)^{\fr{n}{2}}
=y_i\le(\fr{1}{\lm_i\sqrt{\Ompi}}\ri)^{\fr{n}{2}}
\eea
 where we have
approximated  $H_{min}^2\simeq H_i^2=V_i/3y_i^2=\Lm_c^{4+n}\phi_i^{-n}/3y_i^2$
in eq.(\ref{ymin})  since $N_{min}-N_i\ll 1$ and we
have taken
from eq.(\ref{lm1}) $\phi_{min}=n/\lm_{min}\simeq n\sqrt{\Ompi}$ and $\phi_i=n/\lm_i$.
If we assume that the initial value of $\phi_i=n/\lm_i=n\Lm_c$ then
eq.(\ref{ymin}) gives
 \be\la{ymin2}
 y_{min}=y_i\le(\fr{\Lm_c}{\sqrt{\Ompi}}\ri)^{n/2}.
 \ee
We see that $y_{min}=O(\lm_i^{-n/2})\simeq O(\Lm_c^{n/2})\ll y_i$ if
 $\Ompi$ is not too
small.

 Shortly after $y$ reaches its minimum value  the
  scaling period begins. In
this period we neglect the quadratic $x,y \ll 1$ terms  in
eqs.(\ref{eqFRW2}) to find:
\be \la{yNHN}
\frac{y_N}{y}=-\frac{H_N}{N}
\ee
which leads to $yH=H_{min} y_{min}=cte$. Notice
that a constant $Hy$ leads to a constant potential
since $V=3H^2y^2$ and therefore
$\lm$ and $\phi$ will be constant during this
scaling period, i.e.
\be\la{lm2}
\lm (N_{min})\simeq \lm (N_2)
\ee
where we have defined the scale $N_2$ as the end of the
scaling period.
Furthermore, still neglecting the quadratic terms on $x$ and $y$ in the
third equation of system (\ref{eqFRW2}) we
get the expressions
\bea \la{Hscal}
H&=&H_{min} e^{-\frac{3}{2}\gm_b (N-N_{min})}
\nonumber\\
y&=&y_{min} e^{\frac{3}{2}\gm_b (N-N_{min})}.
\eea
We can take in eqs.(\ref{Hscal}) $N_{min}\simeq N_i$ and
$H_{min}\simeq H_i$ as discussed above, but $y_{min} \ll y_i$.

In the same approximation ($x,y \ll 1$) the evolution of
$x$ is given by
\be\la{xn}
x_N=(-3+\fr{3}{2}\gm_b)x
\ee
and we have a decreasing $x$ since $x_N\leq 0$ for all values of $\gm_b$.
The evolution is
\be\la{solx}
x(N)=x_{min}(N_{min})e^{(-3+\fr{3}{2}\gm_b)(N-N_{min})}.
\ee

The scaling period finishes when eq.(\ref{yNHN}) is no longer valid
and the first term in $y_N$ of eqs.(\ref{eqFRW2}) cannot be neglected.
This will happen when $\lm x$ is of order one and $x$ will be of the
same order of $y$, i.e. $\gmp$ will be significant
larger then zero (say $\gmp \sim 0.1$).

At the end of the scaling period we have $1/x_2\sim \lm(N_2)=\lm(N_{min})$ and
\be\la{omp2}
\Omp(N_2)= y^2(N_2)+x^2(N_2)\sim \lm(N_{min})^{-2}\sim \Ompi
\ee
as seen from eq.(\ref{lm1}). The value of $\Omp(N_2)$
depends on the initial $\Ompi$ and can be much smaller than one. This
happens in general for tracker fields since the growing of
$\Omp$ from $\Omp(N_2)\simeq \Ompi \ll 1$
after the end of the scaling period to $\Ompo=0.7$ gives
enough time for  $\gmp$ to grow from $\gmp\sim 0$
to its tracker value $\gmptr=\lm^2\Omp/n^2$.
On the other hand if $\lm(N_2)$ is of the order  one
then $x(N_2)\sim 1$ and $\Omp(N_2)\sim \Ompo=0.7$ and  there
is not enough time to
allow  $\gmp$ to grow to its tracker value and
one has $0<\gmpo\leq \gmptr$.

When the scaling period is over, $\lm$ and the field $\phi$ start to evolve
again to the minimum of $V$. $\gmp$ grows and reaches its tracker value
and may or may not remain constant for long period of time. At the end
 the late time behavior has
$\lm \rightarrow 0$ and $\Omp \sim y^2 \rightarrow 1$ with $\gmp \rightarrow
0$.

\subsubsection{Parameters}\la{para}

There are only four independent
parameters that fix  the cosmological evolution of the models from its initial value
to present day.  These parameters are
$\Ompi$, $\Lm_c, y_{min}$ and the value of $\gmpo$ today.
All other quantities can be derived from them.

The amount
of e-folds between the initial time $N_i$ at $\Lm_c$  and $N_1$,
the scale where $w$ goes from $w=1$ to $w=-1$, is set
by the condition $x\sim y$ and both are mush smaller than one.
We use the evolution of $x,$ from eqs.(\ref{Hscal}) and (\ref{solx})
to get
\be\la{n1ni.a}
N_1-N_i=\fr{1}{3}Log\le[\fr{x_{min}}{y_{min}}\ri]
\ee
were we have assumed $N_i\simeq N_{min}$. Eq.(\ref{n1ni.a})
is independent of $\gm_b$.
We can take $x_{min}=\sqrt{\Ompi}$, $y_i\simeq \sqrt{\Ompi}$
and for an IPL model we have $y_{min}\simeq y_i(\Lm_c/\sqrt{\Ompi})^{n/2}$
and eq.(\ref{n1ni.a}) gives
\be\la{n1ni.b}
\D N_1\equiv N_1-N_i=\fr{n}{6}Log\le[\fr{\sqrt{Ompi}}{\Lm_c}\ri]
\ee
The amount
of e-folds between the initial time $N_i$ at $\Lm_c$
and the end  of the scaling
period $N_2$ is given by eqs.(\ref{Hscal}), (\ref{lm1}) and (\ref{omp2})
with $y=y_2\sim x_2=1/\lm(N_{min})\sim \sqrt{\Ompi}$
 and $y_{min}\sim y_i(\Lm_c/\sqrt{\Ompi})^{n/2}$
 (with $N_i\sim N_{min}$)  giving
\be\la{n2-ni}
N_2-N_i=\fr{2}{3\gmb}Log\le[\fr{y_2}{y_{min}}\ri]
=\fr{n}{3\gmb}Log[\fr{\sqrt{\Ompi}}{\Lm_c}]
\ee
and we have taken in the last eq.(\ref{n2-ni}) $y_i\sim \sqrt{\Ompi}$

Finally, the amount of time between  the end of the scaling regime $N_2$
and present day , which
we denote be $N_o$, with $\Ompo=0.7$ can be approximated from eq.(\ref{Hscal})
by
\be\la{no-n2}
 \D N_o\equiv  N_o-N_2=\fr{2}{3\gmb}Log\le[\fr{y_o}{y_2}\ri]
=\fr{1}{3}Log\le[\fr{\Ompo}{\Ompi}\ri]
\ee
where we have taken $\Ompo\sim y_o^2$ since $y_o^2 > x_o^2$
and $\gmb=1$ today.

Summing eqs.(\ref{n2-ni}) and (\ref{no-n2}) we have
\be\la{NT}
N_o-N_i=\fr{1}{3\gmb}Log[\Lm_c^{-n}\Ompi^{(n/2-1)}\Ompo]
\ee
which gives the total scale $\Delta N_T\equiv N_o-N_i$  between
the initial time at $\Lm_c$ and present day.

We see that the size of the different regions can
be determine by the four parameters $\Lm_c,\Ompi, y_{min}$ and $\wpo$.

\subsection{Evolution of $\gmp=\wp+1$}\la{evol.gmp}

We have seen the evolution of $x,y,H$ in the preceding
subsection and we would like now to show  how $\wp$ evolves
in a general framework. The tracker solution is
just a special case (or the late time evolution)
of the general behavior of the scalar field shown here.

The evolution of the equation of state parameter, $\gmp=1+\wp$, as
given by eq.(\ref{gmp}) has a generic behavior for all scalar
fields independent of its potential. We see that $(\gmp)_N=0$ has
three solutions, $\gmp=2,0$ and $\lm^2\Omp/3$ (or
$\wp=1,-1$ and $\lm^2\Omp/3-1$).

The parameter $\gmp$ will be
most of the time in either of the three critical points.
Independent of its initial value it will go quite rapidly
to $\gmp=1$ and remain there for a long period of time.
The fast increase in $\gmp$ is because $\lm_i \gg 1$.
Afterwards  it will sharply go
to $\gmp=0$ and stay there during almost the same amount of time
as in the first stage. Finally it will go to its
last period given by the tracker value
$\gmptr=\lm^2\Omp/3$ where it will remain.

The first stage ($\gmp=2$) represents a scalar field
whose kinetic energy density dominates ($E_k\gg V$), it is called
the kinetic region, and the energy density redshifts as $\rp \sim
a^{-6}=e^{-6N}$. The second period ($\gmp=0$) is valid when the potential energy
dominates ($E_k\ll V$) and, therefore, the field remains constant
and the energy density redshifts as a cosmological constant with
$\rp \sim a^{0}\sim cte$. The last critical value gives a
${\gmptr}= \lm^2\Omp/3$ and this is called the tracker value and it
is not completely constant. The
energy density redshifts as a tracker field $\rp\sim a^{-3\gmptr}=e^{-3N\gmptr}$.
We will denote the beginning
of the kinetic term by $N_i$ and the end by $N_1$. The
second period ($\gmp=0$) starts at $N_1$ and finishes
at $N_2$.

In table \ref{gmp} we summarize the behavior of $\rp, \Omp, \gmp$
for the different cases.

Let us define the quantity
\be\la{A}
A\equiv \lm \sqrt{\fr{\Omp}{3\gmp}}.
\ee
We see from  eq.(\ref{gmp}) that the sign of $\gm_{\phi N}$
depends if $A>1$ or $A<1$.
For $A>1$ we have $\gm_{\phi N} \geq 0$ and the value $\gm_{\phi
max}=2$ or $\wp=1$, which is the maximum value for $\gmp$,  is a
stable point, i.e. as long as $A > 1$ the parameter $\gmp$ will
grow towards its maximum value and will stay at this point.
 For $A<1$ we have $\gm_{\phi N} \leq 0$
and the value $\gmp=\wp+1=0$ will be a stable point also.

\subsubsection*{First Period, $\gmp=\wp+1=2$}\la{first}

In
this first period one has $\gmp=2$ and
the redshift of $\phi$ is much faster then radiation
or matter and  $\Omp$ will decrease.  We will have at
the end of the period $N=N_1$,  $\gmp\sim 2, \lm=O(1)$ and
 $\Omp(N_1)=r_1/(1+r_1)$ with
\be\la{r1}
r_1\equiv \fr{\rp (N_1)}{\rb (N_1)}=\fr{\rp (N_i)}{\rb (N_i)}\;\;
 e^{-3(N_1-N_i)(2-\gmgm)}.
\ee

At the initial time since $\lm_i \gg 1$ we have $A> \lm  \sqrt{\Ompi/6}\gg1$
 since
$\gmp \leq 2 $. From eq.(\ref{gmp}) we see that the derivative
of $\gmp \gg 1$  and $\gmp$
will rapidly go to its maximum value 2.

The kinetic period must stop
at some point since $A$ which is proportional to $\Omp$ will
decrease as well and it will eventually become less than one and the
sign of $\gm_{\phi N}$ will become negative.
The value of $A$ at the beginning of the scaling
regime (which is when $y$ reaches its minimum value)
is $A=\lm_{min}\sqrt{\Omp(N_{min})/3 \gmp(N_{min})}
=[\gmgm+\Ompi(2-\gmgm)]/2$
(c.f. eq.(\ref{lm1})) which is already smaller than one. However,
even though $A<1$ and $(\gmp)_N<0$, the period of $\gmp\simeq 2$ remains
valid
for a long period of time since
 $\gmp(N_{min})=2x_{min}^2/(x_{min}^2+y_{min}^2)
\simeq 2(1-y_{min}^2/x_{min}^2)\ll 1$ for $x_{min}^2=\Ompi \gg y_{min}^2$.
So we expect $\gmp$ to be close to two for a long period of time
and it will drop to
one only  until $y \sim x$. This
can be quite large depending on the value of $y_{min}$ , i.e.
it depends on $\Lm_c$ and the initial conditions $y_i$ since
$y_{min}=O(y_i\lm_i^{-n/2})\simeq O(y_i\Lm_c^{n/2})$ as can be seen
from eq.(\ref{ymin}). The evolution of $y$ in the scaling period
is given by eq.(\ref{Hscal}).

How many e-folds $N_1-N_i$
has this period depends on the initial conditions and on the
phase transition scale $\Lm_c$.

\subsubsection*{Second Period, $\gmp=\wp+1=0$}\la{second}

The second stage starts when $x\sim y$. We are still in the scaling regime
with $yH=cte$ and since we have $\Omp \ll 1$, as seen from eq.(\ref{ompi}),
$A \ll 1 $. The quantity $\gmp$ has been decreasing and it will arrive
at its minimum possible value $\gmp=0$ or $\wp=-1$. As long as
$A<1$ the value of $\gmp\sim 0$ will remain constant and the $\phi$ will be
constant during all this time, this is the second part of
the scaling regime. The transition time between $\gmp=1.9$ and $\gmp=0.1$
is quite  short, about $\Delta N=1$, so we do not
need to take into account the transition period.

Notice that even
though $\gmp \sim 0$ it is not completely zero since
$A^2=\lm^2\Omp/3\gmp<1$ implies that $0<\Omp<3\gmp/\lm^2$.  Since in this
period $\rp$ redshifts much slower than radiation or matter, $\Omp$
will start to increase and $A$ will eventually become larger then
one again. This is the end of period two. During all this period we
have, $\gmp\sim 0, \phi \sim cte, \lm=\lm_{min}$ and the evolution of
$\Omp(N_2)=r_2/(1+r_2)$ is given by
\be\la{r2}
r_2\equiv \fr{\rp (N_2)}{\rb (N_2)}=\fr{\rp (N_1)}{\rb (N_1)}\;\;
 e^{3(N_2-N_1)\gmgm}.
\ee

Since during this period the field $\phi $ remains constant
the value of $\lm(N_2) \sim \lm(N_1)\sim \lm(N_{min}) $
is also constant.

The second period ends (as the scaling period) when eq.(\ref{yNHN})
is no longer valid and the first term in the equation $y_N$ of
eqs.(\ref{eqFRW2}) cannot be neglected. This happens
for $x(N_2)\sim \lm(N_2)^{-1}$ (c.f. discussion below eq.(\ref{Hscal})).

\subsubsection*{Third Period, $\gmp=\wp+1=\lm^2\Omp/3$}\la{third}

The third period starts when $\gmp$ is not too small
(i.e. $x$ is comparable to $y$ and $\gmp=O(1/10)$. During
all this time we have  $A>1$ again. However, in this case it will
$\gmp$ will not arrive at the maximum value
$\gmp=2$ since $\lm$ is not large and it will stabilize at
\be\la{gtr}
\gmptr=\lm^2\Omp/3.
\ee
and we will have $\Omp(N_o)=r_3/(1+r_3)$ with
\be\la{r3}
r_3\equiv \fr{\rp (N_o)}{\rb (N_o)}=\fr{\rp (N_2)}{\rb (N_2)} \;\;e^{-3(N_o-N_2)
(\gmptr-\gmgm)}
\ee
 If $\gmptr < \gmgm$ then $\Omp \rightarrow 1$.
 While
$\lm^2\Omp$ remains constant we have the constant
tracker value for $\gmp$ or $\wp$. A constant $\gmp$  is
possible when $\Omp \ll 1$. However, at late
times the attractor value will be  $\gmptr \rightarrow 0$
and $\Omp \rightarrow 1$ since $\Omp$ is constrained to be
smaller than one and $\lm \rightarrow 0$. But, even
for $\gmptr$ not constant the evolution of
$\gmptr$ in eq.(\ref{gtr}) is valid and the value
generalizes the tracker behavior.

For an inverse power law potential $V=V_o\phi^{-n}$ we
have $\lm=n/\phi$ and $\gmptr=n^2\Omp/3\phi^2$ which
is the valued obtained by \ci{tracker},\ci{chris2}.

\section{Duration of the Periods}\la{duration}

In order to know the relative size of the
different periods  we
can use eqs.(\ref{r1}) and (\ref{r2}). Let us
define $\Delta N_1=N_1-N_i$ and $\Delta N_2=N_2-N_1$,
they give the amount of e-folds during the $\gmp=2$
and $\gmp=0$ periods, respectively. Combining both
eqs.(\ref{r1}) and (\ref{r2}) we have
\be\la{r12}
\fr{r_2}{r_i}=\fr{\rp (N_2)\rb (N_i)}{\rb (N_2)\rp (N_i)}
=e^{-3\Delta N_1(2-\gmgm)+3
\Delta N_2 \gmgm}
\ee
If the exponent in eq.(\ref{r12}) is zero  than we
will have $r_2=r_i$ which implies that $\Omp(N_2)=\Ompi(N_i)$.
Solving for $\Delta N_2$ in eq.(\ref{r12}) we
obtain
\be\la{N1}
\Delta N_2=\Delta N_1\le(\fr{2}{\gmb}-1\ri)+
\fr{1}{3\gmb}Log\le[\fr{r_2}{r_i}\ri]
\ee
If we  use the
result of quintessence evolution at the beginning
and end of the scaling period $\Omp(N_2)=\Ompi(N_i)$
given in eq.(\ref{omp2}) we have $r_2=r_i$.
For  matter, $\gmb=1$, and  eq.(\ref{N1}) gives
$\Delta N_2=\Delta N_1  $ while
for radiation, $\gmb=4/3$, and
$\Delta N_2=\Delta N_1/2 $.

The universe has been dominated by matter for
a period of $N_o-N_{rm}\simeq 8$, where $N_o$ stands
for present day value and $N_{rm}$ for the scale
at radiation-matter equivalence.

Including the third period we have from
eqs.(\ref{r1}), (\ref{r2}) and (\ref{r3})
\be\la{r13}
\fr{r_3}{r_i}=\fr{\rp (N_o)\rb (N_i)}{\rb (N_o)\rp (N_i)}
=e^{-3\Delta N_1(2-\gmgm (N_1))+3
\Delta N_2 \gmgm (N_2) -3\Delta N_o (\gmptr -1)}=\fr{r_2}{r_i}\;\;
e^{-3\Delta N_o (\gmptr -1)}
\ee
where we have assumed that the third period is
already at the matter dominated epoch, $\gmb(N_o)=1$.
It is clear from eq.(\ref{r13}) that the size $\Delta N_o$
and the value of $\gmptr$ will set
the initial energy density $r_i,\; \Omp(N_i)=r_i/(1+r_i)$
assuming that the final stage
of period three is  today $N_o$ and $\Omp(N_o)=\Ompo=0.7$ which gives $r_3=7/3$.
If we take in eq.(\ref{r13}) the equality $\Ompi(N_i)=\Omp(N_2)$
which implies $r_i=r_2$, then we can express
$r_i=r_3e^{3\Delta N_o(\gmptr -1)}$
and $\Ompi=r_i/(1+r_i)$.
Of course, on the other hand, if we know $\Ompi$ then we can determine
$\Delta N_o \gmptr$.

As a  function of
$\Delta N_T=\Delta N_2+\Delta N_2+\Delta N_o$  we can estimate
the magnitude of the phase transition scale
$\Lm_c$. From $\Lm_c\equiv V_i^{1/4}=(3H^2_iy^2_i)^{1/4}$ and
using  the approximation that $\Omp \ll 1$ during almost all
the time between present day and initial time (at $\Lm_c$)
we have
\be\la{hi}
H_i=H_o \;e^{3\gmb \Delta N_T/2}
\ee
giving a scale
\be\la{Lmc}
\Lm_c=3^{1/4} \sqrt{y_i H_o}\;e^{3\gmb\Delta N_T/4}.
\ee
The scale $\Lm_c$ increases with larger  $\Delta N_T$.
From eqs.(\ref{NT}) and (\ref{Lmc}) we can derive the
order of magnitude for $\Lm_c$ in terms of $n$ and $H_o$
giving $\Lm_c=H_o^{2/(4+n)}$. If we know $\D N_T$ then
we can determine $\Lm_c$ and the power $n$ for IPL models.

How long do the periods last depends on the  models
and by varying the size of $\Delta N_2,\Delta N_o$
and $\gmptr$ we cover all models.

If $\Delta N_o=0$ and $\Delta N_2 > \Delta N_{rm}=N_o-N_{rm}$  then
the model would be undistinguishable from a
true cosmological constant $\gmp=\wp+1=0$ since during
all the matter domination era the equation of state would
be $\gmp=0$. If we have $\Delta N_o >\Delta N_{rm}$
then the model reduces to tracker models with a
constant $\gmpo$ during all the matter domination era.
So, our model contains the tracker and cosmological constant
as limiting cases.

More interesting is to see if we can determine the
nature and scale of the dark energy. For this to happen
a late time phase transition must take place
such that $\Lm_c$ is at $\Delta N_T   \lesssim  \Delta N_{rm}$.

\section{Analysis of CMB spectra}\la{anal}

We will now analyze the generic behavior of a fluid with
equation of state divided in four different regions
with $w=1/3,1,-1,\wtr$. We will vary the sizes of
the regions and we will determine the effect
of having regions with $w=1/3$ or $w=1$ in contrast
to a cosmological constant or a tracker field (with $-1<\wtr=cte<-2/3$).
This analysis is valid for
all kinds of fluids with the specific equation of state
and it is also the generic behavior of scalar fields.
We will compare to the  model $w_{tr}=-0.82$ which was
found to be a better fit to CMB than a true
cosmological constant \ci{neww}.

\subsection{Effect of Radiation Period, $w=1/3$}

The first section we have $w=1/3$ and the fluid (scalar field)
redshifts as radiation. As long as the fluid has $w=1/3$
its energy density will remain the same compared to radiation.
If during nucleosynthesis the fluid has $w=1/3$ then
the BBNS bound requires the $\Omp(NS)<0.1-0.2$ \ci{NS,NSC}.

\begin{figure}[htp!]
\begin{center}
\includegraphics[width=14cm]{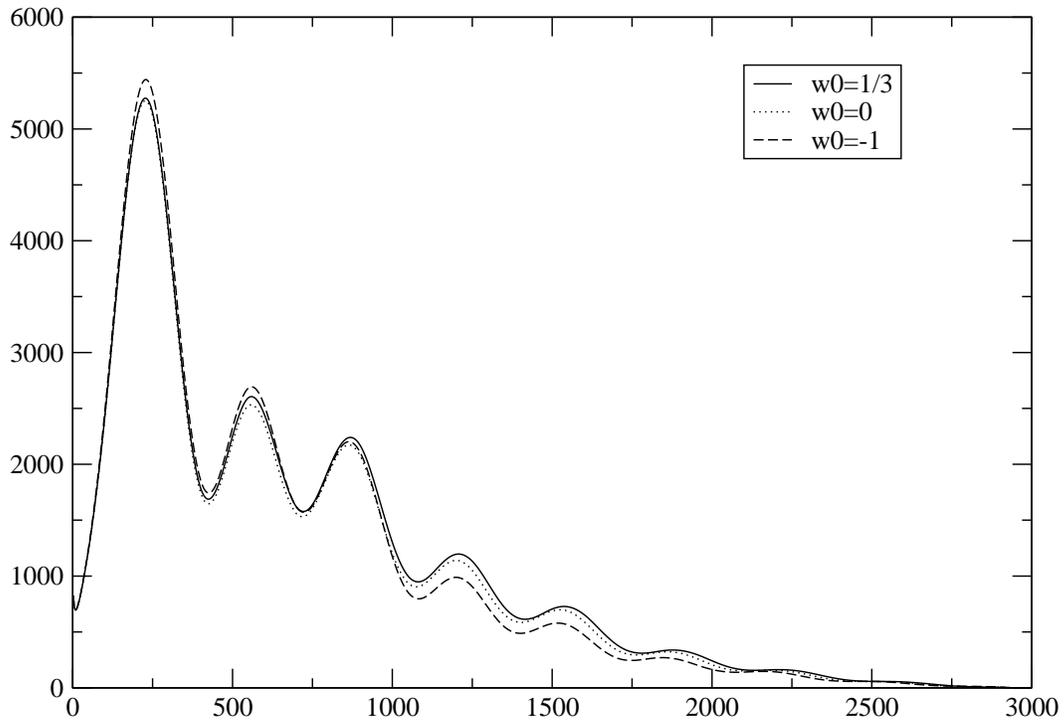}
\end{center}
\caption{\small{We show the effect of the radiation on the CMB
era for $N<N_i$ by changing $w=1/3,0,-1 $ with $\D N_T=2\D N_2=9$.
}}
\la{f.w1}
\end{figure}

In fig.(\ref{f.w1}) we show the different CMB for
$w=1/3,0,-1$ for $\Delta N_T=N_o-N_i=9, \Delta N_2=N_2-N_1=4.5$ and
$\D N_o=N_o-N_2=0$.
We have chosen $\D N_T=9$ because it is the smallest
value satisfying the condition $\D N_1=\D N_2,\D N_o=0$
and giving the correct CMB spectrum. We have  taken
$w=1,-1$ for $N_i<N<N_1$ and $N_1<N<N_2=N_o$, respectively.

We see that
the first and second peaks are suppressed for $w=1/3$ compared to
$w=-1$ while the third peak is enhanced. The positions of
the first two peaks is basically the same and the position of
the third peak is moved from 868 to 864 ($0.4\%$), for $w=1/3,-1$
respectively. For smaller $N_i$, i.e. more distant from present day,
the effect is suppressed. It is not surprising since the
$N_i$ would be further way from energy-matter equality and its
effect on CMB would be less important.

\subsection{Effect of First Period, $w=1$}

\begin{figure}[htp!]
\begin{center}
\includegraphics[width=14cm]{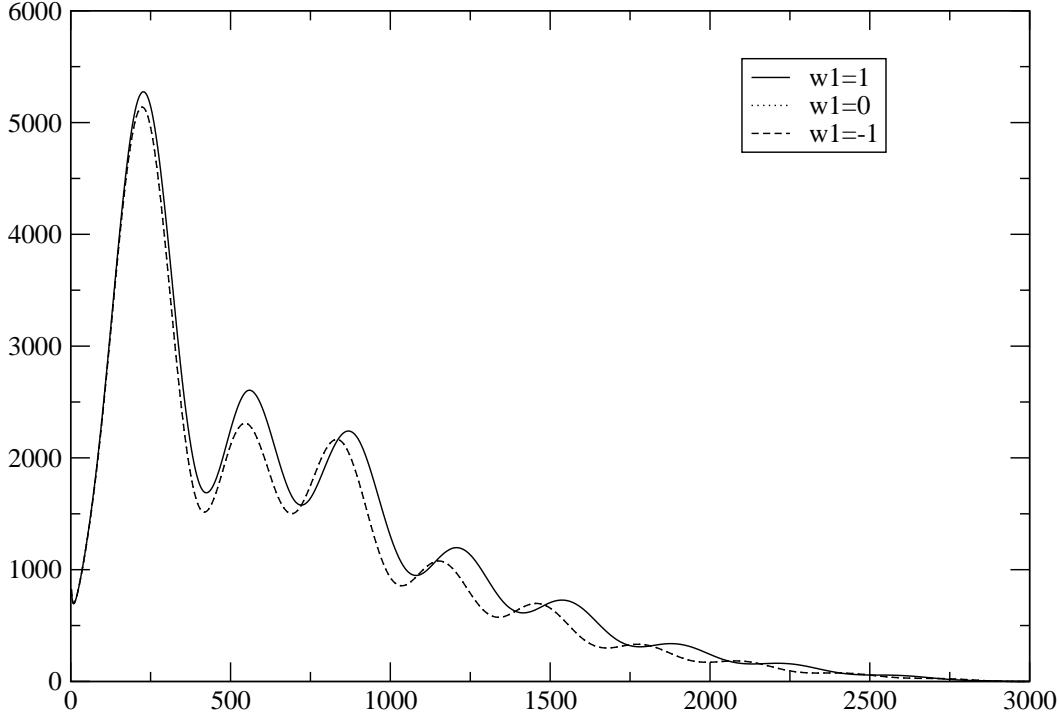}
\end{center}
\caption{\small{ We show  the effect of the kinetic era
with $N_i<N<N_1$ by varying  $w=1,0,-1$ and $\D N_T=2\D N_2=9$.
The vertical axes is $l(l+1)c_l/2\pi (\mu K^2)$.
}}
\la{f.w2}
\end{figure}

In fig.(\ref{f.w2}) we show the CMB for different values
of $w=1,0,-1$ during  $N_i<N<N_1$ and take $w=1/3$
for $N< N_i$ while $w=-1$ for $N_1<N_2=N_o$.
The effect of having a kinetic period enhances
the first three peaks and shifts the
spectrum to higher modes, i.e. higher $l$. The
curve for $w=0$ is indistinguishable from the $w=-1$
one. The position and hight of the peaks
are $p_1=(227,5275),p_2=(559,2605),p_3=(868,2240)$
for $w=1$ while for $w=-1$ we have
$p_1=(224,5138),p_2=(545,2310),p_3=(832,2165)$
giving a percentage difference
$p_1=(1.3\%,2.6\%),p_2=(2.5\%,12.7\%),p_3=(4.3\%,3.4\%)$.
We see that the largest discrepancy is the
altitude of the second peak.

The difference in hight and positions may
in principle distinguish between a cosmological
constant and a scalar field, or any fluid
with the specific equation of state behavior.

The effect of having $w\neq -1$ during
$N_i<N<N_1$   is not
significant for  $w=0$ and would be even less
for a tracker fields with  $\wtr=cte<-2/3$
but it is observable for $w=1/3$.

\subsection{Equal length periods, $\D N_1\equiv N_1-N_i=N_2-N_1\equiv \D N_2,\;
 \D N_o\equiv N_o-N_2=0$}

We have studied the case with $\D N_1=\D N_2,\;\D N_o=0$.
In fig.(\ref{f.wn1=n2})
we show the behavior for different values
of $\D N_T=2\D N_2=6,8,9,12,16$.

There is a lower limit of $\D N_T$ that gives an acceptable  CMB spectrum.
The lower limit is $\D N_T \geq 9$.
For smaller $\D N_T$ the peaks move to
the right of the spectrum and the hight increases
giving a spectrum not consistent with the CMB data.

For larger  $\D N_T>9$ the curves tend to
the cosmological constant. It is not surprising since for
large $\D N_T=2 \D N_2$ it means
that we have a larger time with $w=-1$ and in the case
that  $\D N_2>N_o- N_{rm}$
the universe content, after matter radiation equality, would have been
given by matter and a fluid with $w=-1$, i.e. a cosmological
constant.

\begin{figure}[htp!]
\begin{center}
\includegraphics[width=14cm]{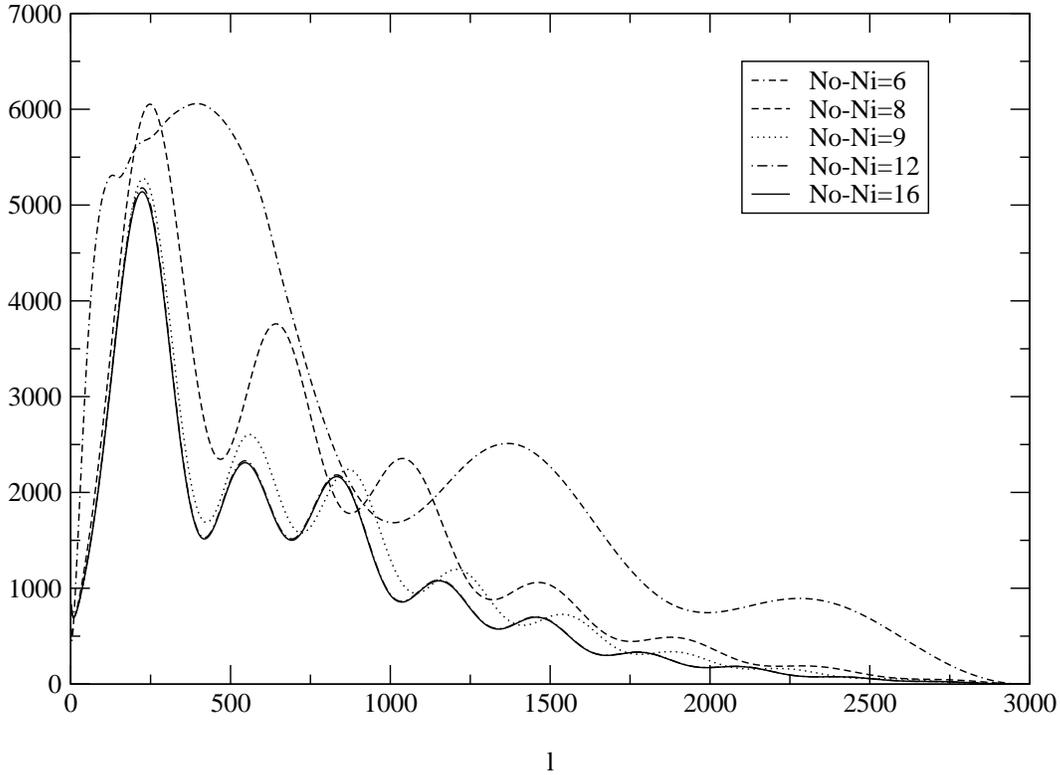}
\end{center}
\caption{\small{We show the effect on the CMB
by varying the $\D N_1=\D N_2=3,4,4.5,6.8$  with $\D N_o=0$.
The vertical axes is $l(l+1)c_l/2\pi (\mu K^2)$.
}}
\la{f.wn1=n2}
\end{figure}

\subsection{Scaling condition $\Omp(N_i)=\Omp(N_2)=0.1$}

Following the discussion in sec.(\ref{evol.xyH}), we now that
a scalar field will end up its scaling period with
a $\Omp$ equal to its starting value, i.e. $\Omp(N_i)=\Omp(N_2)=0.1$.
We have taken this value of $\Omp$ since for $N>N_i$ the
energy density behaved as radiation and
we have to impose
the   nucleosynthesis
bound on relativistic degrees of freedom $\Omp(NS)\leq 0.1-0.2$.
Imposing this condition we have determined the evolution
of the CMB for three different values of $\wtr=-1,-0.82,-0.7$.
We have chosen to analyze the $\wtr=-0.82$ because it was found to be
the best fit tracker model by \ci{neww}. We
have $w=1/3$ for $ N\leq N_i$, $w=1$ for $ N_i\leq N \leq N_1$,
 $w=-1$ for $ N_1\leq
N\leq N_2$ and $w=\wtr$ for $ N_2\leq N\leq N_o$. The value of
$N_2$ is determined so that the energy density grows
from $\Omp(N_2)=0.1$ to   $\Omp=0.7$ today.
This conditions sets the range  of the period to
$N_o-N_2=1.03, 1.25, 1.47$ for
$w=-1,-0.82,-0.7$ respectively.

\begin{figure}[htp!]
\begin{center}
\includegraphics[width=14cm]{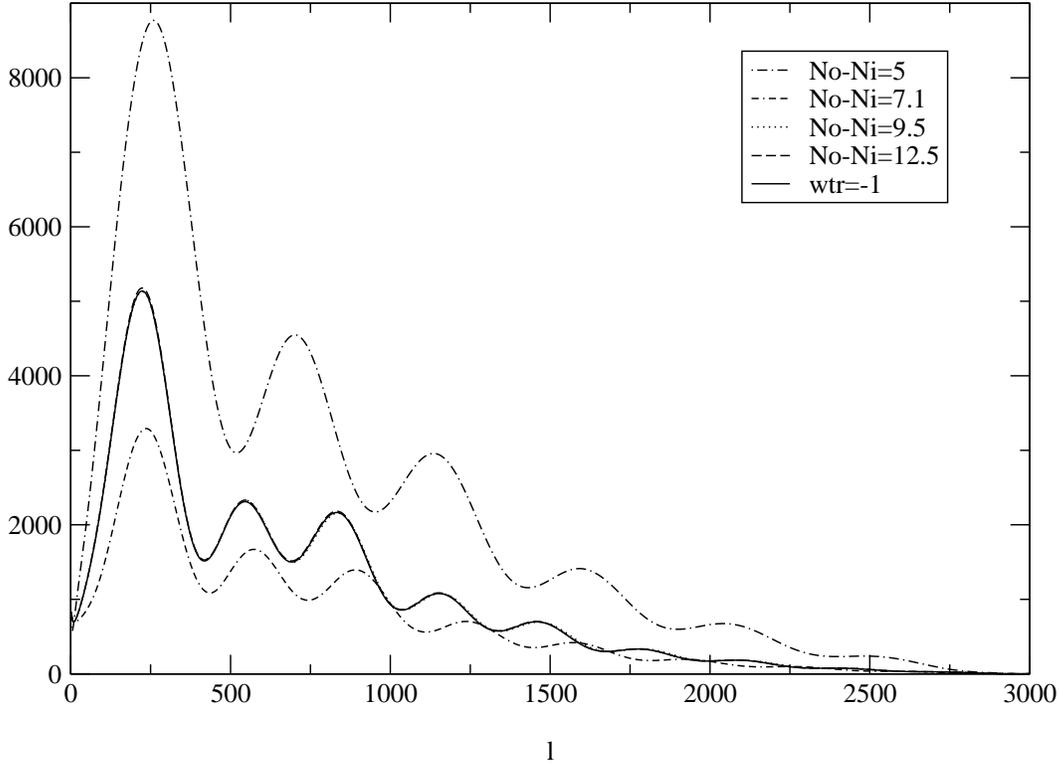}
\end{center}
\caption{\small{We show the effect on the CMB
by varying the $\D N_T=N_o-N_i=5,7.1,9.5,12.5$, with
the constrain $\Ompi(N_i)=\Omp(N_2)=0.1$
and $\wptr=-1, \D N_2=1.03$ and we include
the cosmological constant $w\equiv -1$,
for comparison.
The vertical axes is $l(l+1)c_l/2\pi (\mu K^2)$.
}}
\la{f.w=1}
\end{figure}

\begin{figure}[htp!]
\begin{center}
\includegraphics[width=14cm]{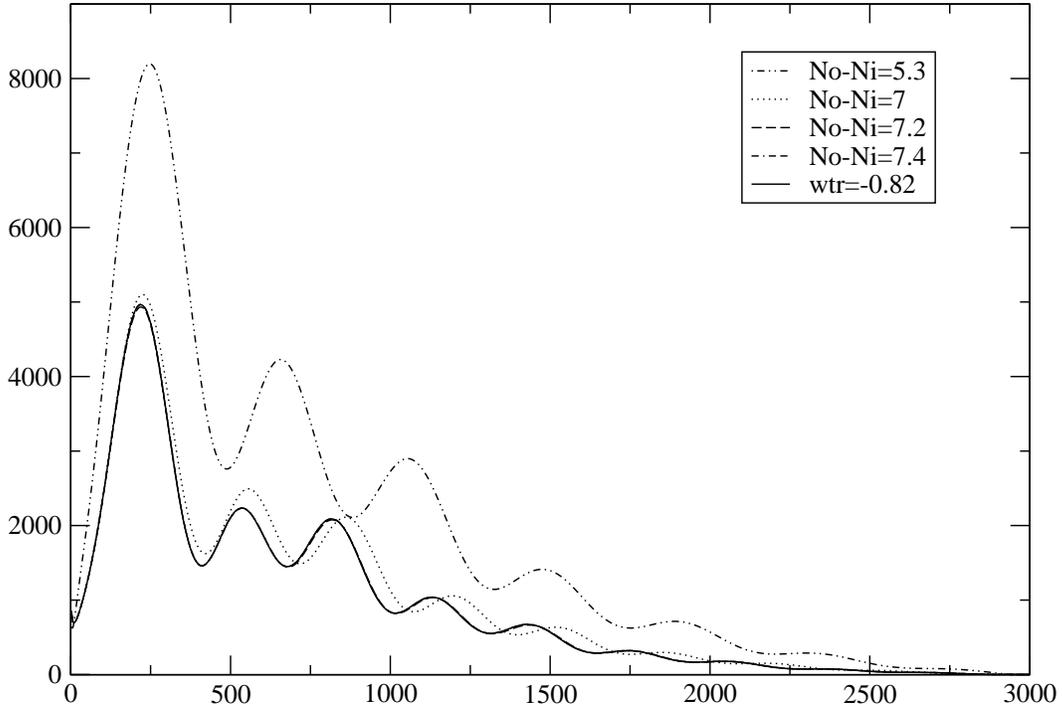}
\end{center}
\caption{\small{We show the effect on the CMB
by varying the $\D N_T=N_o-N_i=5.3,7,7.2,7.4$, with
the constrain $\Ompi(N_i)=\Omp(N_2)=0.1$
and $\wptr=-0.82, \D N_2=1.25$. We also include
the tracker with constant $w\equiv =0.82$
The vertical axes is $l(l+1)c_l/2\pi (\mu K^2)$.
}}
\la{f.w=82}
\end{figure}

\begin{figure}[htp!]
\begin{center}
\includegraphics[width=14cm]{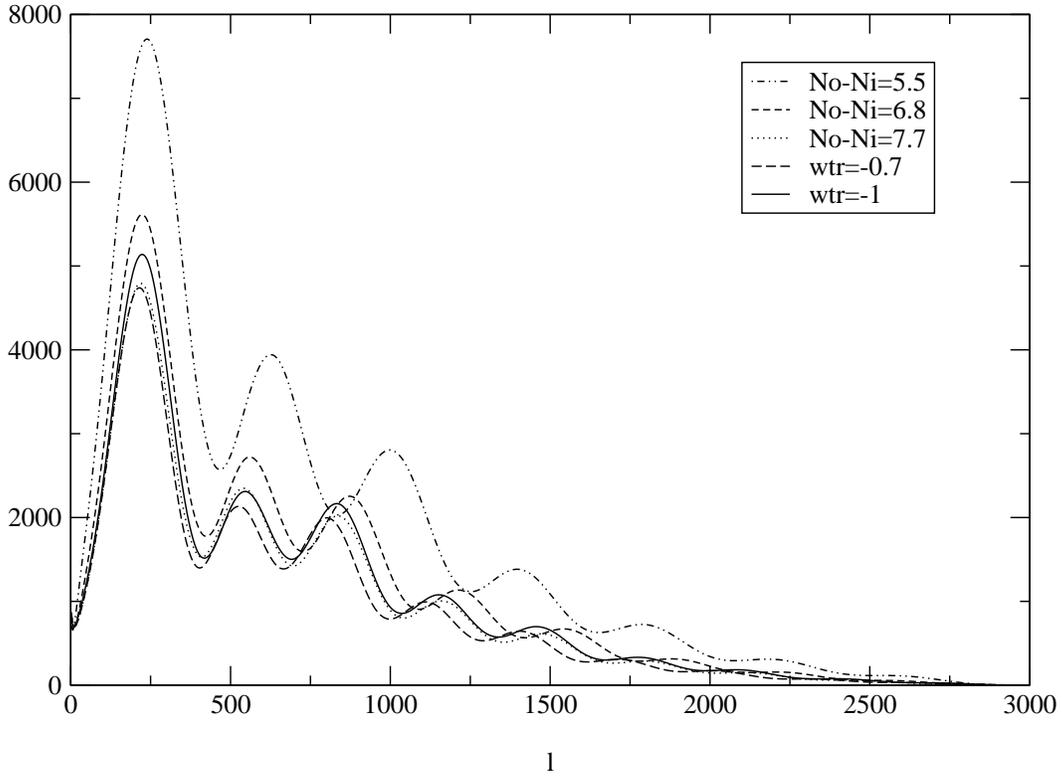}
\end{center}
\caption{\small{We show the effect on the CMB
by varying the $\D N_T=N_o-N_i=5.5,6.8,7.7$, with
the constrain $\Ompi(N_i)=\Omp(N_2)=0.1$
and $\wptr=-0.7, \D N_2=1.47$. We also include
the tracker $w=-0.7$ and the constant $w=-1$.
The vertical axes is $l(l+1)c_l/2\pi (\mu K^2)$.
}}
\la{f.w=7}
\end{figure}

In fig.(\ref{f.w=1}),(\ref{f.w=82}) and (\ref{f.w=7}) we show
the curves for different  values of $N_i$ with the restriction
that $\Omp(N_i)=\Omp(N_2)=0.1$ and for $\wp=-1,-0.82,-0.7$,
respectively.

In the case of
$w=-1$ we have $\D N_o=1.03$ and
that the smallest acceptable model has $\D N_T=8.5$,  $\D N_2=3.6$,
see fig.(\ref{f.w=1}).
The best model has $\D N_T=8.88, N_1-N_o=3.7$ and peaks
$p_1=(224,5133),p_2=(549,2363),p_3=(840,2178)$.

For $w=-0.82$, fig.(\ref{f.w=82}), the best fit for tracker models, we have
$\D N_o=N_o-N_2=1.25$ and the minimum acceptable distance is $\D N_T=N_i-N_o=7.19,
\D N_2=3.9$. Smaller values of $\D N_T$ give a spectrum with peaks
too large and second and third peaks moved to the right
(high $l$ modes). For large $\D N_T$ the spectrum tends to the
tracker spectrum $w=-0.82$.

The best
model has   $\D N_T=N_1-N_o=7.6,\D N_2=3.1$ with peaks
and position $p_1=(223,5016),p_2=(551,2450),p_3=(851,2102)$.
We have compared the $\chi^2$ of the models and
the $\D N_T=7.6, \wp=-0.82$ model has a
better fit than the tracker model with constant $w_{tr}=-0.82$,
which was found to be the best tracker fit \ci{neww}. We
see that having a dynamical $w$, is not only more
reasonable from a theoretical point of view but it
fits the data  better.

Finally, we consider $w=-0.7$ for $N>N_2$. In this case
we have $\D N_o=N_o-N_2=1.47$ and
the minimum acceptable model has $\D N_T=6.8, \D N_2=3.6$,
while the best model has
$\D N_T=7.3, \D N_2=3.8$  with peaks
$p_1=(222,4954),p_2=(550,2422),p_3=(853,2035)$.

We see that in all three cases $w=-1,-0.82,-0.7$, with
condition $\Omp(N_i)=\Omp(N_2)=0.1$ we have
a minimum acceptable value of $\D N_T$ and
for smaller $\D N_T$ the peaks  move to the
right of the spectrum and the hight of the peaks
increases considerably. This conclusion is generic
and sets a lower limit to $\D N_T$, the
distance to the phase transition
scale $\Lm_c$, or equivalently it sets a lower
limit to $\Lm_c$.

The smallest  $\D N_T$ is set by the largest acceptable
$\wpo$ (here we have taken it to be $\wpo=-0.7$)
giving in our case a $\D N_T=6.8$ for $\Ompi=0.1$.
This result puts a constraint on how late the phase
transition can take place. In terms of the energy
$\Lm_c=\rho_{\phi i}^{1/4}=[\Ompi 3H^2_i]^{1/4}$ we can set a lower
value for  the transition scale.
Using eq.(\ref{Lmc}) with  $\Ompi=0.1$ and$\D N_T=6.82$ we get
\be\la{ph.tr}
\Lm_c=\rho_{\phi i}^{1/4}=2\times 10^{-10} GeV=0.2 eV
\ee
i.e. for models with a phase transition below
eq.(\ref{ph.tr}) the CMB will not agree with the
observations. This result is independent
of the type of potential.

Furthermore, we now that for inverse power potential
there is un upper limit to $\Lm_c$ coming by requiring
that $\wpo<-2/3$. The limiting value assuming
$\Ompi\leq 0.1$, for $V=\Lm^{4+n}\phi^{-n}$, is
 $n<1.8$ giving  $\Lm_c=4\times10^{-3} GeV\simeq H_o^{2/(4+n)}$.
Therefore, for  IPL potentials the only acceptable models have
phase transition scale
\be
4 \times 10^{-3} GeV > \Lm_c > 2\times 10^{-10} GeV.
\ee

\section{Conclusions}\la{concl}

We have analyzed the CMB spectra for a fluid with an equation of state
that takes different values. The values are $w=1/3,1,-1,\wtr$ for
$N$ in the regions
$N_i-N_{Planck}, N_1-N_i, N_2-N_1, N_o-N_2$, respectively.
The results are independent
of the type of fluid we have. The
cosmological constant and the tracker models
are special cases of our general set up.

We have shown that the evolution of a scalar field,
for any potential that leads to an accelerating universe
at late times, has exactly the kind of behavior described above.
It starts at the condensation scale $\Lm_c$ and enters a period
with $w=1$, then it undergoes a period with $w=-1$
and finally ends up in a region with $-1\leq w \leq -2/3$. We
have shown that  the  energy density at the end
of the scaling period (end of $w=-1$ region) has the same
energy ratio as in the beginning, i.e.
$\Omp(N_i)=\Omp(N_2)$. The time it spends on the last region
depends on the value of $\Omp(N_2)$ and on $w$ during this time.
Before the phase transition scale $\Lm_c$ we are assuming that all
particles were at thermal equilibrium and  massless in the quintessence
sector. At the phase transition scale $\Lm_c$ the particles
acquire a mass and a non trivial potential.

We have shown that models with  $w=1/3,1,-1,\wtr$ have a better
fit to the data then tracker or cosmological constant. Furthermore,
we have determined the effect of the first two periods
$w=1/3$ and $w=1$ and even though the effect is small
it is nonetheless observable.

In general, the CMB spectrum sets a lower limit
to $\D N_T$, which implies a lower limit
to the phase transition scale $\Lm_c$. For smaller
$\D N_T$ the CMB peaks are moved to the right of the spectrum
and the hight increases considerably.

For any $\Ompi$  the CMB sets a lower limit to the phase transition
scale. In the case of
 $\Ompi(N_i)= 0.1$ the limit
is $\Lm_c= 0.2 eV$ for any  scalar potential. We
do not take $\Ompi$ much larger because we should
comply with the NS bound on relativistic degrees
of freedom $\Ompi\leq 0.1-0.2$. If we take
$\Ompi \ll 0.1$ then the constrain on the phase transition scale
will be less stringent since the effect of the scalar
field is only relevant recently ($\Omp \ll 1$
during all the time before present time).

For inverse power law potentials we can also set un upper limit
to $\Lm_c$ and for $\Ompi\leq 0.1$ it gives an inverse
power $n\leq 1.8$ and $\Lm_c\leq 4\times 10^{-3} GeV$.
In this class of potentials only models with
$4\times 10^{-3} GeV> \Lm_c > 2\times 10^{-8} GeV$
would give the correct $\wpo$ and CMB spectrum.

Acknowledgments

We would like to thank useful discussions
with C. Terrero. This work was supported in part by
 CONACYT project 32415-E and
DGAPA, UNAM project IN-110200.

\thebibliography{}

\footnotesize{

\bib{CMB} {P. de Bernardis {\it et al}. Nature, (London) 404, (2000)
955, S. Hannany {\it et al}.,Astrophys.J.545 (2000) L1-L4}

\bib{SN1a} {A.G. Riess {\it et al.}, Astron. J. 116 (1998) 1009; S.
Perlmutter {\it et al}, ApJ 517 (1999) 565; P.M. Garnavich {\it et
al}, Ap.J 509 (1998) 74.}

\bib{structure}{ G. Efstathiou, S. Maddox and W. Sutherland,
 Nature 348 (1990) 705.
  J. Primack and A. Klypin, Nucl. Phys. Proc. Suppl. 51 B, (1996),
30}

\bib{gen.dark}P.S. Corasaniti astro-ph/0210257;P.S. Corasaniti, E.J. Copeland
astro-ph/0205544,S. Hannestad, E. Mortsell Phys.Rev.D66:063508,2002;
J. P. Kneller, G. Steigman astro-ph/0210500

\bib{w}{S. Perlmutter, M. Turner and  M. J. White,
Phys.Rev.Lett.83:670-673, 1999; T. Saini, S. Raychaudhury, V. Sahni
and  A.A. Starobinsky, Phys.Rev.Lett.85:1162-1165,2000 }

\bib{neww} Carlo Baccigalupi, Amedeo Balbi, Sabino Matarrese,
Francesca Perrotta, Nicola Vittorio, Phys.Rev. D65 (2002) 063520

\bib{wette} Michael Doran, Matthew J. Lilley, Jan Schwindt,
 Christof Wetterich, astro-ph/0012139; Michael Doran, Matthew Lilley,
  Christof Wetterich  astro-ph/0105457

\bib{wcop}  P.S. Corasaniti, E.J. Copeland, Phys.Rev.D65:043004,2002

\bib{tracker} I. Zlatev, L. Wang and P.J. Steinhardt, Phys. Rev.
Lett.82 (1999) 8960;  Phys. Rev. D59 (1999)123504

\bib{chris1}{A. de la Macorra and C. Stephan-Otto, Phys.Rev.Lett.87:271301,2001 }

\bib{generic}  A.R. Liddle and R.J. Scherrer, Phys.Rev.
D59,  (1999)023509

\bib{mio.scalar}{A. de la Macorra and G. Piccinelli, Phys.
Rev.D61 (2000) 123503}

\bib{liddle}E.J. Copeland, A. Liddle and D. Wands, Phys. Rev. D57 (1998) 4686

\bib{brax}  P. Brax, J. Martin, Phys.Rev.D61:103502,2000

\bib{r=m}{E.Kolb and M.S Turner,The Early Universe, Edit. Addison Wesley 1990}

\bib{ax.mod} A. de la Macorra hep-ph/0111292

\bib{NS} {K. Freese, F.C. Adams, J.A. Frieman and E. Mottola, Nucl. Phys. B 287
(1987) 797; M. Birkel and S. Sarkar, Astropart. Phys. 6 (1997)
197.}

\bib{NSC} {C. Wetterich, Nucl. Phys. B302 (1988) 302, R.H. Cyburt, B.D. Fields,   K. A. Olive, Astropart.Phys.17:87-100,2002  }

\bib{ax.coinci}{A. de la Macorra, Int.J.Mod.Phys.D9 (2000) 661 }

\bib{chris2}{ A. de la Macorra and C. Stephan-Otto, Phys.Rev.D65:083520,2002 }

\end{document}